\documentclass{PoS}
\usepackage{graphicx}
\usepackage{epstopdf}
\usepackage{amsmath}

\title{Evolution of higher moments of multiplicity distribution}

\ShortTitle{Evolution of higher moments of multiplicity distribution}

\author{\speaker{Radka Sochorov\'a}\\
        \v{C}esk\'e vysok\'e u\v{c}en\'i technick\'e v Praze, FJFI, B\v{r}ehov\'a 7, 115 19 Praha 1, Czech Republic\\
        E-mail: \email{sochorad@fjfi.cvut.cz}}

\author{Boris Tom\'a\v{s}ik\\
        Univerzita Mateja Bela, Tajovsk\'eho 40, 97401 Bansk\'a Bystrica, Slovakia\\
        \v{C}esk\'e vysok\'e u\v{c}en\'i technick\'e v Praze, FJFI, B\v{r}ehov\'a 7, 115 19 Praha 1, Czech Republic\\
        E-mail: \email{boris.tomasik@fjfi.cvut.cz}}

\abstract{Evolution of a multiplicity distribution can be described with the help of master equation.
	We first look at 3rd and 4th factorial moments of multiplicity distributions and derive their equilibrium values. From them 
	central moments and other ratios can be calculated. 
	We study the master equation for a fixed temperature, because we want to know how fast different moments of the multiplicity distribution 
	approach their equilibrium value. Then we investigate the situation in which the temperature of the system decreases. 
	We find out that in the non-equilibrium state, higher factorial moments differ more from their equilibrium values than the lower moments 
	and that the behaviour of a combination of the central moments depends on the combination we choose.}

\FullConference{XIII Quark Confinement and the Hadron Spectrum - Confinement2018\\
		31 July - 6 August 2018\\
		Maynooth University, Ireland}

\begin{document}

\section{Motivation}

The main motivation of this work is the observation that overall observed multiplicity of different types of particles from ultrarelativistic 
heavy-ion collisions agrees with the statistical model at temperatures above $160~\mathrm{MeV}$ \cite{Andronic:2017pug}. 
On the other hand, the phase transition temperature can be also determined from multiplicity fluctuations. Higher-order 
susceptibilities determined by lattice QCD are then compared with data on higher-order moments of the multiplicity distribution. 
The extracted temperature is usually lower than $160~\mathrm{MeV}$ \cite{Alba:2014eba}.

The main aim of this work is to study the evolution of the multiplicity distribution in fireball that cools down after chemical freeze-out. 
We want to know the answer to the question: Can different moments of the multiplicity distribution be influenced by the temperature decrease differently? 
This is important, because then we could obtain different apparent temperatures from different moments. 

For the description of the evolution of the multiplicity distribution we use a master equation. 
Particularly, we focus on the higher factorial moments from which all other kinds of moments, 
e.g.\ central moments or the coefficients of skewness and kurtosis, can be calculated. 
We first study in Section~\ref{s:2} the relaxation of factorial moments when the temperature is fixed. 
Then, in Section~\ref{s:3} the cooling scenario is investigated. In Section \ref{s:4} we draw conclusions 
on the extracted apparent temperature and we summarise in Section~\ref{s:5}.


\section{Relaxation of factorial moments}
\label{s:2}

We will consider a binary reversible process $a_1 a_2 \leftrightarrow b_1 b_2$. Here, none of the involved 
species are identical to each other and it is understood that $b$ particles carry a conserved charge while the $a$ particles do not.
We will also assume that there is a sufficiently large pool of $a's$ which is basically untouched by this chemical process. For our study 
we shall investigate multiplicity distribution of species that conserve an abelian charge, e.g. strangeness.

The master equation \cite{1} for $P_n(\tau)$, the probability of finding $n$ pairs $b_1 b_2$, formulated in dimensionless time 
$\tau$ has the following form
\begin{equation}
\frac{dP_n(\tau)}{d \tau} = \epsilon \left[ P_{n-1}(\tau) - P_n(\tau)\right] - \left[ n^2 P_n(\tau) - \left( n+1\right) ^2 P_{n+1}(\tau)\right],
\end{equation} 
where $n$ goes from 0 to $\infty$, and the constant $\epsilon$ is defined as 
$\epsilon = G \left\langle N_{a_1} \right\rangle \left\langle N_{a_2} \right\rangle /L $. 
Here, $ \left\langle N_{a_1} \right\rangle $,  $ \left\langle N_{a_2} \right\rangle $ are (initial) averaged number of particles $a_1, a_2 $, 
and $G, L$ stand for the momentum-averaged cross-section of the gain process ($a_1 a_2 \rightarrow b_1 b_2$) and the loss process ($b_1 b_2 \rightarrow a_1 a_2$), respectively
\begin{equation}
\label{eq:1}
	G  =  \langle \sigma_G v \rangle \, , \qquad
	L  =  \langle \sigma_L v \rangle\,  .
\end{equation}
The dimensionless time $\tau = t/\tau_0^c$ is formulated with the help of the relaxation time  
$\tau_{0} ^c = V/L$, where $V$ is the effective volume.

From the master equation we can derive the equilibrium distribution of the factorial moments. 
For this purpose, the master equation can be converted into a partial differential equation for the generating function \cite{1}
\begin{equation}\label{eq:2}
g(x, \tau) = \sum_{n=0}^{\infty} x^n P_n (\tau),
\end{equation}
where $x$ is an auxiliary variable. If we multiply (\ref{eq:2}) by $x^n$ and sum over $n$, we find that \cite{1} 
\begin{equation}
\frac{\partial g(x, \tau)}{\partial \tau}= (1-x)(xg''+g'-\epsilon g),
\end{equation}
where $g' = \partial g/\partial x$. The generating function obeys the normalisation condition
\begin{equation}
g(1, \tau) = \sum_{n=0}^{\infty} P_n (\tau) = 1.
\end{equation}
The equilibrium solution, $g_{eq} (x)$, must not depend on time, thus it obeys the following equation
\begin{equation}
x g_{eq} ^{''} + g_{eq} ^{'} - \epsilon g_{eq} = 0.
\end{equation} 
The solution that is regular at $x = 0$ is then given by
\begin{equation}
g_{eq} (x) = \frac{I_0 (2 \sqrt{\epsilon x})}{I_0 (2 \sqrt{\epsilon})}.
\end{equation}
Here, $I_0(x)$ is the Bessel function. From the derivatives of the generating function we can easily determine the 
factorial moments, since $\langle N!/(N-k)!\rangle = \partial^kg/\partial x^k$. 
(The first and second factorial moments have been  calculated 
in \cite{1,2}.) 
	\begin{eqnarray}\label{eq:3}
	\langle N \rangle_{eq} 
	& = & \sqrt{\varepsilon} 
	\frac{I_1(2\sqrt{\varepsilon})}{I_0(2\sqrt{\varepsilon})} \\
	\langle N(N-1) \rangle_{eq} 
	& = & - \frac{1}{2} \sqrt{\varepsilon}
	\frac{I_1(2\sqrt{\varepsilon})}{I_0(2\sqrt{\varepsilon})}  + \frac{1}{2}\varepsilon 
	\frac{I_2(2\sqrt{\varepsilon}) + I_0(2\sqrt{\varepsilon})}{I_1(2\sqrt{\varepsilon})} \\
        \left \langle \frac{ N!}{(N-3)!} \right \rangle_{eq}
        & = & 
	\frac{3}{4} \sqrt{\varepsilon}\frac{I_1(2\sqrt{\varepsilon})}{I_0(2\sqrt{\varepsilon})} \nonumber - \frac{3}{4} \varepsilon\left ( 1 +  \frac{ I_2(2\sqrt{\varepsilon})}{I_0(2\sqrt{\varepsilon})}\right ) \\
	&&+ \frac{1}{4} \varepsilon^{3/2} 
	\frac{I_3(2\sqrt{\varepsilon}) + 3I_1(2\sqrt{\varepsilon})}{I_0(2\sqrt{\varepsilon})}\\
	 \left \langle \frac{ N!}{(N-4)!} \right \rangle_{eq}
	 & = & -\frac{15}{8} \sqrt{\varepsilon} \frac{I_1(2\sqrt{\varepsilon})}{I_0(2\sqrt{\varepsilon})} \nonumber + \frac{15}{8} \varepsilon \left (\frac{I_2(2\sqrt{\varepsilon})}{I_0(2\sqrt{\varepsilon})} +1\right) \\
	&&-\frac{3}{4} \varepsilon^{3/2} 
	\frac{3I_1(2\sqrt{\varepsilon}) +I_3(2\sqrt{\varepsilon})}{I_0(2\sqrt{\varepsilon})} 
 + \frac{1}{8} \varepsilon^2 \left ( 3 + 
	\frac{4I_2(2\sqrt{\varepsilon}) + I_4(2\sqrt{\varepsilon})}{I_0(2\sqrt{\varepsilon})}\right ) \, .
	\end{eqnarray}
	
Now, we let the distribution of the multiplicities relax with the help of master equation. 
We calculate the evolution of \emph{scaled} factorial moments, which are defined as 
\begin{eqnarray}\label{eq:4}
F_2 & = & \frac{\langle N (N-1) \rangle}{\langle N \rangle^2}\\
F_3 & = & \frac{\left \langle N (N-1) (N-2)\right \rangle}{\langle N \rangle^3}\\
F_4 & = & \frac{\left \langle N (N-1) (N-2) (N-3)\right \rangle}{\langle N \rangle^4} \,   .
\end{eqnarray}

For numerical calculations, binomial initial conditions  were used 
\begin{eqnarray}\label{eq:5}
P_0(\tau = 0) & = & 1-N_0\\
P_1(\tau = 0) & = & N_0\\
P_n(\tau = 0) & = & 0\, \qquad \mbox{for}\phantom{m}n>1\, ,
\end{eqnarray}
where $N_0 = 0.005$ $(N_0 = \left\langle N \right\rangle (\tau = 0))$.

\begin{figure}[h]
\centerline{\includegraphics[width=0.7\textwidth]{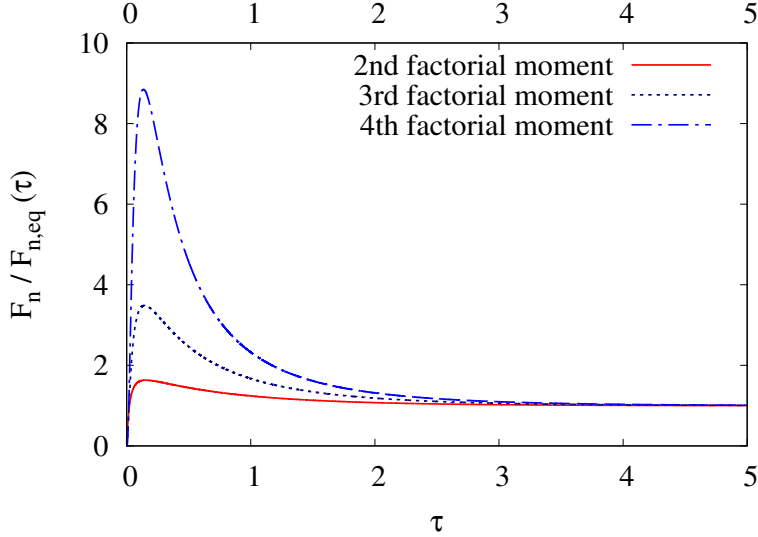}}
	\caption{Time evolution of scaled factorial moments divided by their equilibrium values for
	constant temperature and $\epsilon = 0.1$.}
	\label{fig:1}
\end{figure}

The evolution of second to fourth scaled factorial moments divided by their equilibrium values is shown in Figure \ref{fig:1}.
The value of the parameter $\epsilon$ has been set to 0.1. It is important to say here that we have obtained qualitatively similar 
results also with other sets of parameters. In Fig.~\ref{fig:1} we can see that relaxation time for all moments is the same. 
However, during relaxation higher moments depart further from their equilibrium values than the lower moments. 


\section{Higher moments in a cooling fireball}
\label{s:3}

Realistic description of the fireball evolution must include decreasing temperature. 
If temperature changes, also the relaxation time will change. 
Thus one cannot use the dimensionless time because relaxation time was the typical scale in introducing the dimensionless time. 
Now we need to go back to real time in the master equation and calculate the creation and annihilation terms for each temperature. 
The master equation takes the form
\begin{equation}\label{eq:6}
\frac{dP_n(t)}{dt} = \frac{G}{V} \left\langle N_{a_1} \right\rangle \left\langle N_{a_2} \right\rangle \left[ P_{n-1} (t) - P_n (t) \right] - 
\frac{L}{V} \left[ n^2 P_n (t) - (n+1)^2 P_{n+1} (t) \right] .
\end{equation}
We shall study, how higher moments evolve in a scenario with a decreasing temperature. In our simulations we shall assume that the system is established in equilibrium at the hadronisation temperature $T = 165~\mathrm{MeV}$. The fireball then cools down further. We investigate, 
how does the distribution of multiplicities change.

To answer this question we have used a simple toy model in which the temperature, volume and density behave like in 1D longitudinally boost-invariant expansion (Bjorken scenario). The effective volume grows linearly 
\begin{equation}\label{eq:7}
V(t) = V_0 \frac{t}{t_0},
\end{equation}
the temperature drops according to 
\begin{equation}\label{eq:8}
T^3 (t)= T_0^3 \frac{t_0}{t}
\end{equation}
and the particle density drops like
\begin{equation}\label{eq:9}
\rho (t) = \rho_0 \frac{t_0}{t}.
\end{equation}

In the calculations, we have set $V_0 = 125 ~\mathrm{fm^3}$, $T_0 = 165 ~\mathrm{MeV}$ and $\rho_0 = 0.08 ~\mathrm{fm^{-3}}$
for the initial state of the evolution. Motivated by the femtoscopic measurements we set the final time to $10 ~\mathrm{fm/c}$
and the final temperature to 100~MeV. This  leads then to $t_0 = 2.2 ~\mathrm{fm/c}$. 

For this calculation we have to choose the particular inelastic process. We have chosen the reaction system $\pi^+ + n \leftrightarrow K^+ + \Lambda^0$. For the moment we shall use a parametrisation of the cross-section \cite{3}
\begin{equation}\label{eq:10}
\sigma_{\pi N}^{\Lambda K} = \left \{ 
\begin{array}{lc}
0\, \mbox{fm}^2 & \sqrt{s} < \sqrt{s_0}\\
\frac{0.054 (\sqrt{s} - \sqrt{s_0})}{0.091}\, \mbox{fm}^2 &  \sqrt{s_0}\le\sqrt{s}<\sqrt{s_0}+0.09\,\mbox{GeV}\\
\frac{0.0045}{\sqrt{s} - \sqrt{s_0}} \, \mbox{fm}^2 & \sqrt{s} \ge \sqrt{s_0}+0.09\,\mbox{GeV}
\end{array}
\right .
\end{equation}
where $\sqrt{s_0}$ is the threshold energy of the reaction and the energies are given in $\mathrm{GeV}$. 
Since we will assume density-dependent mass of $\Lambda$, the threshold energy will also depend on the density. 

We shall assume that the mass of $\Lambda$ hyperon depends on baryon density as 
\begin{equation}\label{eq:11}
m(\rho) = -2.2 ~\mathrm{GeV \cdot fm^3} \cdot \rho + m_{\Lambda0}\,  . 
\end{equation}
Hence, the hyperon mass becomes identical to that of the proton at the highest baryon density $\rho_0$ at which our calculations starts, and returns to the vacuum value $m_{\Lambda 0}$ if baryon density vanishes.

\begin{figure}[h]
	\centering
	\includegraphics[width=0.7\textwidth]{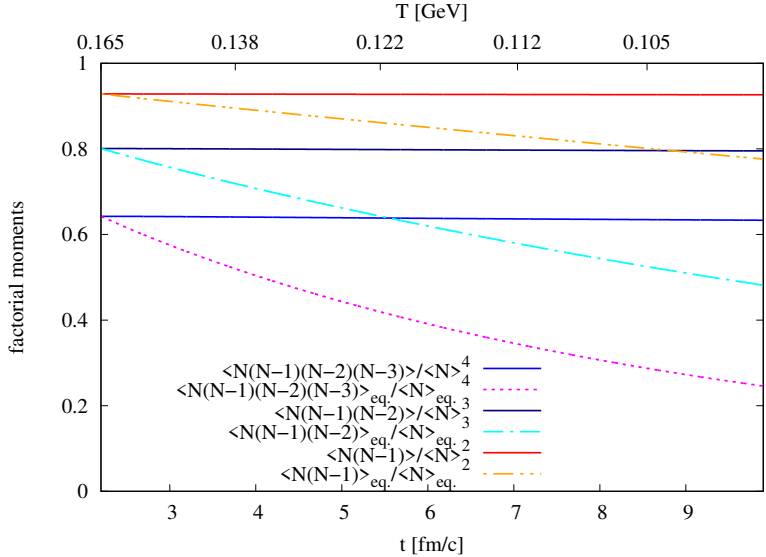}
	\caption{Scaled factorial moments for the gradual change of temperature. 
	Solid lines: evolution of moments according to master equation. Dashed lines: equilibrium values at the given temperature.}
	\label{fig:2}
\end{figure}

The scaled factorial moments for the cooling scenario are shown in Figure~\ref{fig:2}.

\section{The freeze-out temperature}
\label{s:4}

\begin{figure}[t]
	\centering
	\includegraphics[width=0.7\textwidth]{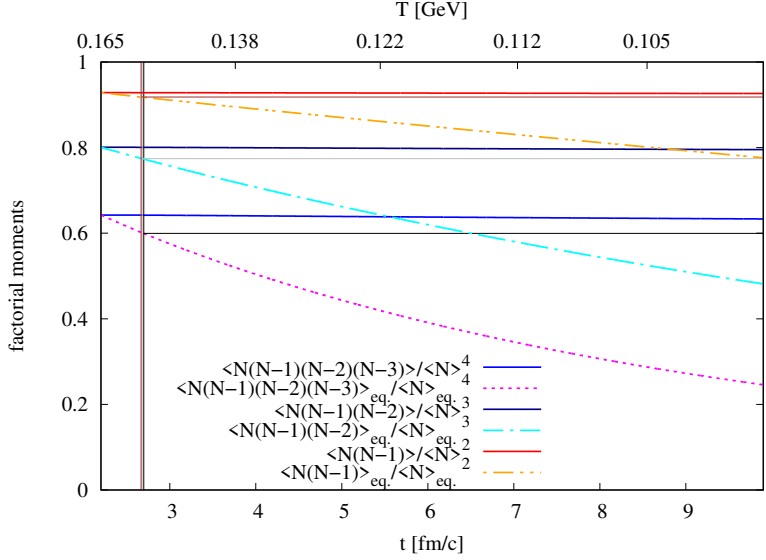}
	\caption{Apparent freeze-out temperature of factorial moments. Solid lines: evolution of moments according to master equation. Dashed lines: equilibrium values at the given temperature.}
	\label{fig:3}
\end{figure}

In Figure~\ref{fig:3} we demonstrate the potential danger in case of extraction of the (apparent) freeze-out temperature from the different moments. 
At the hadronisation temperature we set the moments to equilibrium values, then we let them evolve. 
Let us assume that the evolution is finished at $T = 100$~MeV. All moments are off-equilibrium, there. Auxiliary lines in Fig.~\ref{fig:3}
demonstrate, how different values of the temperature would be obtained from different orders of the moments if they are
interpreted as equilibrated. 

In experimental data, more conveniently, the central moments are used.
\begin{eqnarray}\label{eq:12}
\mu_1 & = & \langle N \rangle = M\\
\mu_2 & = & \langle N^2 \rangle - \langle N\rangle^2 = \sigma^2\\
\mu_3 & = & \langle (N-\langle N\rangle )^3\rangle \\
\mu_4 & = & \langle (N-\langle N\rangle )^4\rangle.
\end{eqnarray}
Often, one uses their combinations like the coefficient of skewness
\begin{equation}\label{eq:13}
S = \frac{\mu_3}{\mu_2^{3/2}} 
\end{equation}
or the coefficient of kurtosis
\begin{equation}\label{eq:14}
\kappa = \frac{\mu_4}{\mu_2^2} - 3.
\end{equation}

We also look at the volume-independent ratios which are often measured. These are, e.g.\ 
\begin{eqnarray}
R_{32} & = & \frac{\mu_3}{\mu_2} = S\sigma\\
R_{42} & = & \frac{\mu_4}{\mu_2} - 3\mu_2 = \kappa\sigma^2.
\end{eqnarray}

\begin{figure}[t!]
	\centering
	\includegraphics[width=1\textwidth]{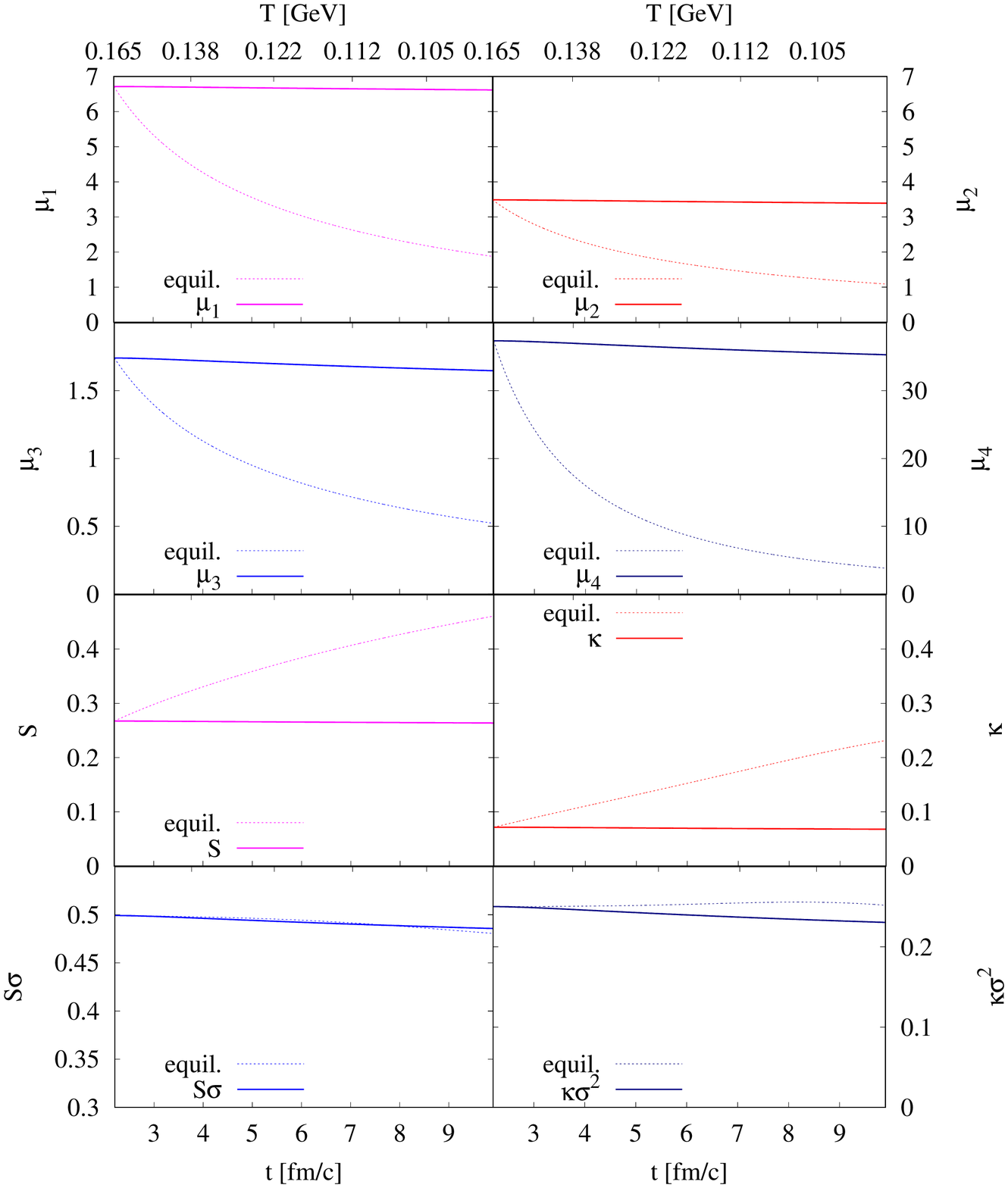}
	\caption{Central moments, skewness, kurtosis and volume-independent ratios $S \sigma$ and $\kappa \sigma^2$ for the scenario with 
	density-dependent mass of $\Lambda$ and the decreasing temperature. Thick solid lines: numerically calculated evolution, thin dotted lines: equilibrium values at the 
	given temperature.}
	\label{fig:4}
\end{figure}

Results are plotted in Fig. \ref{fig:4}.
We can see that while the central moments are decreasing with the time evolution, the coefficients of skewness and kurtosis are increasing. 
Only slight changes are seen for the volume independent ratios $S \sigma$ and $\kappa \sigma^2$. So the extracted apparent temperature 
strongly depends on the chosen observable. In real collisions we have non-equilibrium evolution of the moments and it is very difficult to 
determine the unique freeze-out temperature from them.


\section{Conclusion}
\label{s:5}

If equilibrium is broken, higher factorial moments of the multiplicity distribution depart further from their equilibrium values than the lower 
moments. Evolution of chemical reaction off equilibrium may show different temperatures for different orders of the (factorial or central) 
moments. We demonstrated this on the example of $ \pi^+ + n \leftrightarrow K^+ + \Lambda^0$. The behavior of the combination of the 
central moments depends on the combination of moments we choose. Caution is mandatory when we want to extract the freeze-out 
temperature from higher moments of the multiplicity distributions. 

\begin{acknowledgments}
This work was supported by the grant 17-04505S of the Czech Science Foundation (GA\v{C}R).
\end{acknowledgments}

\end{document}